\documentclass[Crown,sageh,times]{sagej}

\usepackage{moreverb,url}

\usepackage[colorlinks,bookmarksopen,bookmarksnumbered,citecolor=red,urlcolor=red]{hyperref}

\usepackage{soul}
\usepackage[export]{adjustbox}
\usepackage{blindtext}

\newcommand\BibTeX{{\rmfamily B\kern-.05em \textsc{i\kern-.025em b}\kern-.08em
T\kern-.1667em\lower.7ex\hbox{E}\kern-.125emX}}

\def\volumeyear{2022}

\begin{document}

\runninghead{Haridis and Stiny}

\title{Analysis of shape grammars: continuity of rules}

\author{Alexandros Haridis\affilnum{1} and George Stiny\affilnum{1}}

\affiliation{\affilnum{1}Department of Architecture, Massachusetts Institute of Technology, USA}

\corrauth{Corresponding author}

\corrauth{Alexandros Haridis, MIT Architecture-Computation, Room 10-303, 77 Massachusetts Avenue, Cambridge, MA 02139, United States. ORCID iD: \url{https://orcid.org/0000-0003-0369-1428}}

\email{charidis@mit.edu}

\begin{abstract}

The rules in a shape grammar apply in terms of embedding to take advantage of the parts that emerge visually in the appearance of shapes. While the shapes are kept unanalyzed as a computation moves forward, part-structures for shapes can be defined retrospectively by analyzing how the rules were applied. An important outcome of this is that rule continuity is not built-in but it is ``fabricated" retrospectively to analyze the computation as a continuous process. An aspect of continuity analysis that has not been addressed in the literature is how to decide which mapping forms to use to study the continuity of rule applications. This is addressed in this paper using recent results on shape topology and continuous mappings. A characterization is provided that distinguishes the suitable mapping forms from those that are inherently discontinuous or practically inconsequential for continuity analysis. It is also shown that certain intrinsic properties of shape topologies and continuous mappings provide an effective method of computing topologies algorithmically. 

\end{abstract}

\maketitle


\section{Analysis of a computation}

A remarkably surprising and easy to miss aspect of the shape grammar formalism takes center stage in the first publication on rule continuity in Stiny (1994). The continuity of rule applications is not a built-in feature of computations with shapes, but it is defined retrospectively by analyzing how the rules were applied in each state. This approach is certainly counter to the antecedent analysis that is usually required today in computations with digital tools and CAD systems, to ensure that the shapes generated in a computation are always structured into compatible descriptions. But it is really the only natural approach in shape grammars when the rules apply in terms of embedding to take advantage of the parts that emerge visually in the appearance of shapes.

Few publications in the literature have subsequently touched on continuity for rules in shape grammars; in particular, Stiny (1996), Krishnamurti and Stouffs (1997), and Earl (1999). A summary of this topic is also included in Stiny (2006). Continuity analysis involves assigning certain structural descriptions to the shapes that are generated in a computation, in particular finite topologies, such that the descriptions are structurally compatible for consecutive applications of rules. The compatibility of the topologies, that is to say their ``continuity,” depends on the mappings that describe the action of the rules. An aspect of continuity analysis that previous work has not addressed is how to actually go about deciding which mapping forms to use to study the continuity of rules. What is it that makes certain mapping forms suitable and certain other mapping forms unsuitable for continuity analysis? In this paper, we address this question by using some of the recent results on shape topology and continuous mappings published in Haridis (2020a). We characterize the suitable mapping forms by distinguishing them from those that are inherently discontinuous or practically inconsequential for continuity analysis. We characterize the suitable mapping forms by distinguishing them from those that are inherently discontinuous or practically inconsequential for continuity analysis. We additionally show how to compute continuous topologies algorithmically using closure equations similar to those that were initially used in Stiny (1994). The closure equations become more effective by taking advantage of certain intrinsic properties of shape topologies and continuous mappings. The Appendices provide mostly supportive technical material, but they also include some additional results on continuity when shapes have points for basic elements.

\subsection{Continuity is retrospective}

\begin{figure}[t!]
\centering
\includegraphics{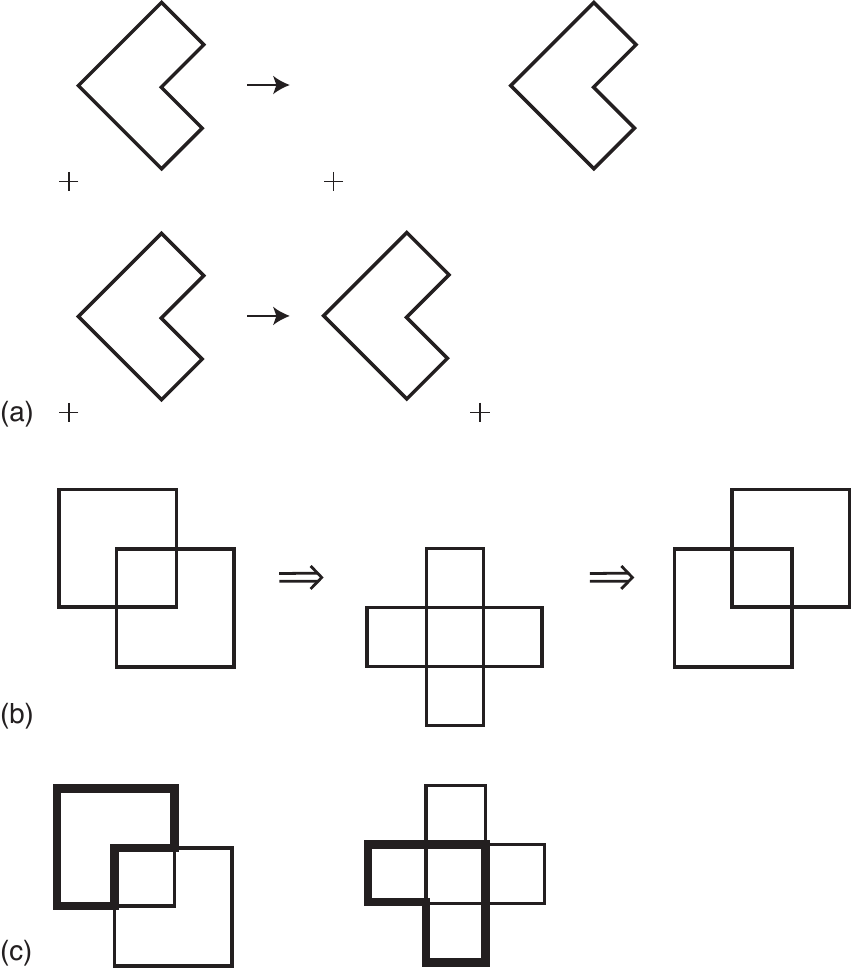}
\label{Figure-1}
\caption{(a) Two rules that translate a chevron right-to-left along its central horizontal axis. (b) A two-step computation that uses the two rules. (c) The parts recognized in the first and second rule applications.}
\end{figure}

\begin{figure}[t!]
\centering
\includegraphics{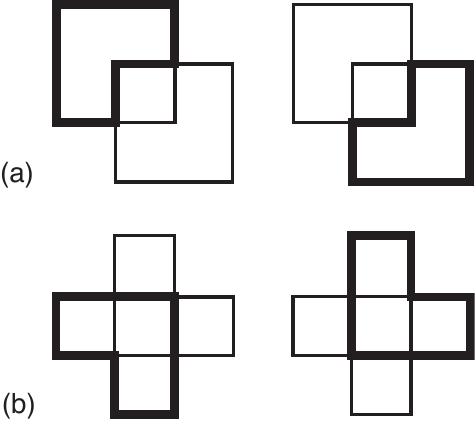}
\label{Figure-2}
\caption{(a) A description of the initial shape in terms of the first rule and (b) a description of the second shape in terms of the second rule. The descriptions are implied in the rule applications in Figure 1(b).}
\end{figure}

Before we expand on the technical side, let us first see how continuity manifests itself in a computation with a simple but instructive example.

The two rules in Figure 1(a) translate a chevron from right to left along its central horizontal axis (the crosshairs in the rules indicate the origin of a standard coordinate system in the Euclidean plane). The rules apply in turn in the two-step computation in Figure 1(b) to produce a shape that is a rotation or a reflection of the initial shape. Even though a rotation or a reflection is nowhere implied in the two rules, the result is achieved without difficulty because the rules apply visually in terms of an embedding relation (the embedding relation for shapes made with maximal line segments is defined in Stiny (1975); see also Stiny, 2006). In particular, the two parts in Figure 1(c) are recognized and changed in the first and second rule applications.

Continuity comes into the picture when we start examining how the rules interact with the parts they alter and what descriptions of shapes are implied by this interaction. Here, by a ``description" for a shape we mean a particular kind of finite structure made up of parts of the shape, namely, a finite topology (see Section \emph{Continuity for rule applications}).

Any shape is represented completely by finite sets of basic elements---line segments in our case---that are maximal with respect to each other. But the basic elements of a shape do not announce conclusively the parts that are visually embedded in the shape. These are instead resolved according to particular rules and the ways in which they are applied in a computation.

\begin{quote}
    "Whenever a rule applies to a shape in a computation, the rule implicitly provides a description of the shape that guides the action of the rule" (Stiny, 1994; s52).
\end{quote}

\noindent The first rule application in Figure 1(b) implicitly provides a description for the initial shape that consists of two discrete chevrons---this is shown in Figure 2(a). The two chevrons correspond to the number of different ways the left side of the first rule can be embedded in the initial shape under transformations.

The second rule application acts similarly. It implicitly provides a description of the second shape that consists of two discrete chevrons that are shown in Figure 2(b). Only now this is not the only description possible with respect to the second rule and the transformations that make its left side part of the cross. There are four different ways of embedding the left side of the rule, as shown in this series

\begin{figure}[h!]
\centering
\includegraphics{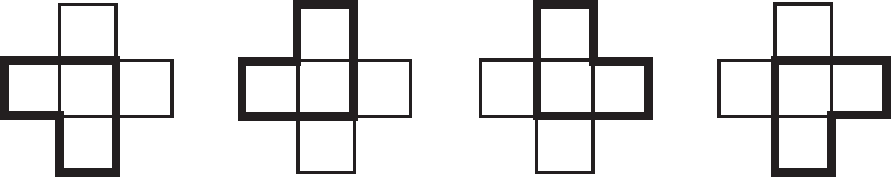}
\label{Figure-1-nocaption}
\end{figure}

\noindent These four emergent chevrons provide alternative descriptions of the cross that are incompatible for the separate applications of the rule. The first of these leads to the final shape in Figure 1(b).

The initial shape in the computation has eight maximal elements. The first rule application implicitly structures the shape with two parts each with six maximal elements, in this way 

\begin{figure}[h!]
\centering
\includegraphics{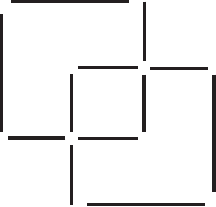}
\label{Figure-2-nocaption}
\end{figure}

\noindent After the first rule application, these divisions in the initial shape fuse so that the cross has eight maximal elements and a new rule can be applied to it under embedding (in shape grammars there is no explicit memory of the prior divisions of shapes in any state of the computation). The second rule application then implicitly structures the cross with two parts each with six maximal elements, in this way

\begin{figure}[h!]
\centering
\includegraphics{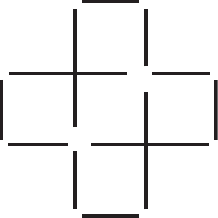}
\label{Figure-3-nocaption}
\end{figure}

\noindent These divisions recognize the emergent chevron, but do they introduce new divisions that are not implied in the divisions of the previous shape? In other words, does the first rule application introduce divisions in the initial shape that anticipate the subsequent recognition of the chevron in the cross? It does not. The divisions of the cross imply that these two highlighted maximal elements of the initial shape

\begin{figure}[h!]
\centering
\includegraphics{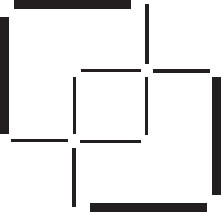}
\label{Figure-4-nocaption}
\end{figure}

\noindent must also be divided to anticipate the chevron in the cross. Without these additional divisions, the two-step computation appears ``discontinuous" in the sense that there is a discernible break or inconsistency in how the separate rule applications structure the shapes.

Descriptive or structural inconsistencies of this kind are not isolated phenomena in shape grammars. They are part and parcel of visual calculating when embedding drives the action of rules. In reality, the need to describe shapes with definite parts or permanent units is unnecessary---in fact, there are cases where such an approach is provably impossible (Jowers et al., 2019). In shape grammars, there are specific mechanisms in place (the reduction rules for basic elements) that keep shapes unanalyzed throughout a computation so that rules can apply uninterrupted to exploit emergent parts in the appearance of shapes.

The above exercise nonetheless illustrates in a direct way the intuition behind continuity analysis: to establish structural descriptions of shapes that appear compatible with how rules are applied in a computation, one must work backward from the results obtained in the computation to the structural descriptions that would yield them. Continuity, or continuous structural change, is not an a priori feature of computations with shapes. It is ``fabricated" retrospectively---after one sees the results of rule applications---in order to explain and analyze a computation as a continuous process. Topology gives a formal framework for investigating questions about continuous structural change in the context of computations with shape grammars. Intuitively, topology gives a synoptic view of how different structural descriptions of shapes interact with one another and it also provides the technical tools to make their interactions continuous. Without topology, or any other kind of structural device for that matter (e.g., parse-trees, graphs, or graphs combining different parse-trees like those in Stiny (2011)), there is no way of investigating questions about structure and structural change in shape grammars---the shapes one deals with in a shape computation remain unanalyzed (i.e., unstructured) throughout all states of the computation. 

To make topology applicable to computations with shape grammars, we need a few key ideas that can allow us to express topology and continuity in terms of finite objects, i.e., shapes made with finitely many maximal elements. A detailed study of finite topology for shapes and in what ways it is similar or different to the topology of infinite objects (e.g., the topology of the real line $\mathbb{R}$ or the real space, $\mathbb{R}^2$ or $\mathbb{R}^3 $) can be found in Haridis (2020a). We will return to this exercise in the section \emph{Examples}, after the necessary technical ideas are in place.

\section{Continuity of rules}

By a \emph{topology} for a shape $S$, we mean a finite set $\mathcal{T}$ of parts of $S$ such that: 

\begin{enumerate}
\item[(1)] The empty shape ($0$) and $S$ itself are in $\mathcal{T}$.

\item[(2)] The sum ($+$) of any number of parts in $\mathcal{T}$ is also in $\mathcal{T}$.

\item[(3)] The product ($\cdot$) of any number of parts in $\mathcal{T}$ is also in $\mathcal{T}$.
\end{enumerate}

\noindent The parts of a shape that are members of a topology are called open parts. The term ``open part" is a primitive concept. The term ``closed part" can be used as the primitive concept, too, without altering significantly the relevant constructions. No matter the terminological choice, what matters most is that any part of a shape that is included in a topology is always equal to its closure.

The way the open parts are related with one another in a topology implies a certain kind of algebraic structure. This algebraic structure is captured by a (finite) lattice: the top element is the shape itself, the bottom element is the empty shape, and sum and product of open parts play the role of join and meet, respectively. The \emph{lattice of open parts} associated with a topology for a shape $S$ is denoted by $OS$. This algebraic construct is important in the following formulation of rule continuity.

Suppose a rule $A \rightarrow B$ is applied to a shape $S$ under a transformation $t$ to generate a new shape $S'$. Let $h: S \rightarrow S^{+}$ be a mapping that describes the rule application $S \Rightarrow S'$, so that $h(S)$ = $S^{+}$ and the shape $S^{+}$ is some part of the shape $S'$. Assume also that $\mathcal{T}$ and $\mathcal{T}'$ are, respectively, the topologies for the shapes $S$ and $S'$. The rule application is continuous if the following two conditions are met:

\begin{enumerate}
\item[(1)] The shape $t(A)$ that the rule recognizes in $S$ is open in $\mathcal{T}$.

\item[(2)] The mapping $h^*: OS^{+} \rightarrow OS$, defined as $h^*(D) = h^{-1}(D)$ for every open part $D$ in $OS^{+}$, is a lattice homomorphism.
\end{enumerate}

\noindent Condition (1) is the same as in Stiny (1994). It is a way of guaranteeing that the part of the shape $S$ that the left side of a rule matches under a transformation is always an open part of $S$. Practically this means that emergent parts of a shape must be recognized, that is, open, in the structure of the shape. Condition (2) tells that the rule $A \rightarrow B$ applies continuously whenever the mapping $h$ that describes its application is continuous for the topologies assigned to the shapes $S$ and $S'$. It is based on the formulation of continuous mappings between shapes given in Haridis (2020a).

The mapping $h^*$ is defined in terms of the preimage operation $h^{-1}$. When $h$ is continuous, it yields a homomorphism from the open parts in $OS^{+}$ to those in $OS$. Thus, for continuity to hold, the topology for $S$ must include all the preimages of the open parts of the topology for $S^{+}$. \hyperref[supA]{Appendix A} provides details on mappings between shapes and the preimage operation.

It is clear from condition (2) that the (preimage) mapping $h^*$ works in the opposite direction of $h$. That is to say, in the opposite direction of the rule application that $h$ describes. Besides having a lot of nice technical features, this formulation also highlights the retrospective character of continuity analysis: the forward action of a rule is analysed backward.

As a homomorphism, the mapping $h^{*}$ preserves all sums and products of open parts in $OS^{+}$:

\vspace{12pt}
\centerline{$h^{*}(D + E) = h^{*}(D) + h^{*}(E)$\;\; and \;\;$h^{*}(D \cdot E) = h^{*}(D) \cdot h^{*}(E)$.} 
\vspace{12pt}

\noindent It follows from (either of) these equations that $h^{*}$ is order-preserving:

\vspace{12pt}
\centerline{If $D \leq E$ in $OS^{+}$, then this implies $h^{*}(D) \leq h^{*}(E)$ in $OS$.} 
\vspace{12pt}

\noindent In other words, $h^{*}$ preserves the embedding order of the open parts determined by the topology assigned to the shape $S^{+}$. ``Smaller" open parts in $OS^{+}$ which are embedded in ``larger" open parts, \emph{continue} to be related in this way after $h^{*}$ maps them backward into $OS$. And, in this way, no open parts recognized in the shape $S^{+}$ imply open parts that are not already included in the topology for the shape $S$.

Some additional properties of the two mappings, $h$ and $h^{*}$, are briefly mentioned. Two or more parts of the shape $S^{+}$ may have the same preimage under $h$. So the backward mapping $h^{*}$ is expected to be many-to-one in the general case. By definition, the forward mapping $h$ maps the shape $S$ into the shape $S^{+} \leq S'$, so that $h(S) = S^{+}$. Thus, $h^{*}(h(S)) = h^{*}(S^{+}) = h^{-1}(S^{+}) = S$ (because $S$ is the largest shape embedded in $S^{+}$ under $h$) and it follows that the mapping $h^{*}$ preserves the top element. However, it does not preserve the bottom element, that is, $h^{*}(0)$ is not necessarily equal to 0 (consider what happens with mappings that erase parts). It is possible to encounter cases where both the top and the bottom elements are preserved, but this depends on the shapes, the topologies, and the mappings involved in an analysis.

\section{How to choose mappings}

Now that we have defined what it means for a rule application to be continuous, let us see in more detail how to choose the mapping $h: S \rightarrow S^{+}$ that describes a rule application.

\subsection{Alternative forms}

The mapping $h$ describes in formal notation how a rule changes the parts of a shape to make another shape. In principle, to define $h$ one must choose the shape $S^{+}$ onto which $h$ maps $S$. One way of doing this, which is implicit in Stiny (1994), is to examine alternative forms of production formulas that fix the rule applications in shape grammars. For example, this formula

\vspace{12pt}
\centerline{$S' = (S - t(A)) + t(B)$} 
\vspace{12pt}

\noindent describes any one-step rule application $S \Rightarrow S'$, where a rule $A \rightarrow B$ applies to a given shape $S$ under a transformation $t$ to generate the new shape $S'$. It has universal applicability, in the sense that it can be used to describe any rule application, for any rule, and for any given shape grammar (in fact, this same formula can be used to uniformly characterize the rule applications in a variety of formal models of computation, including Turing machines, Phrase structure grammars, Post production systems, Markov normal algorithms, etc., as explained in Gips and Stiny (1980)).

\begin{figure}[t!]
\centering
\includegraphics[width=\textwidth]{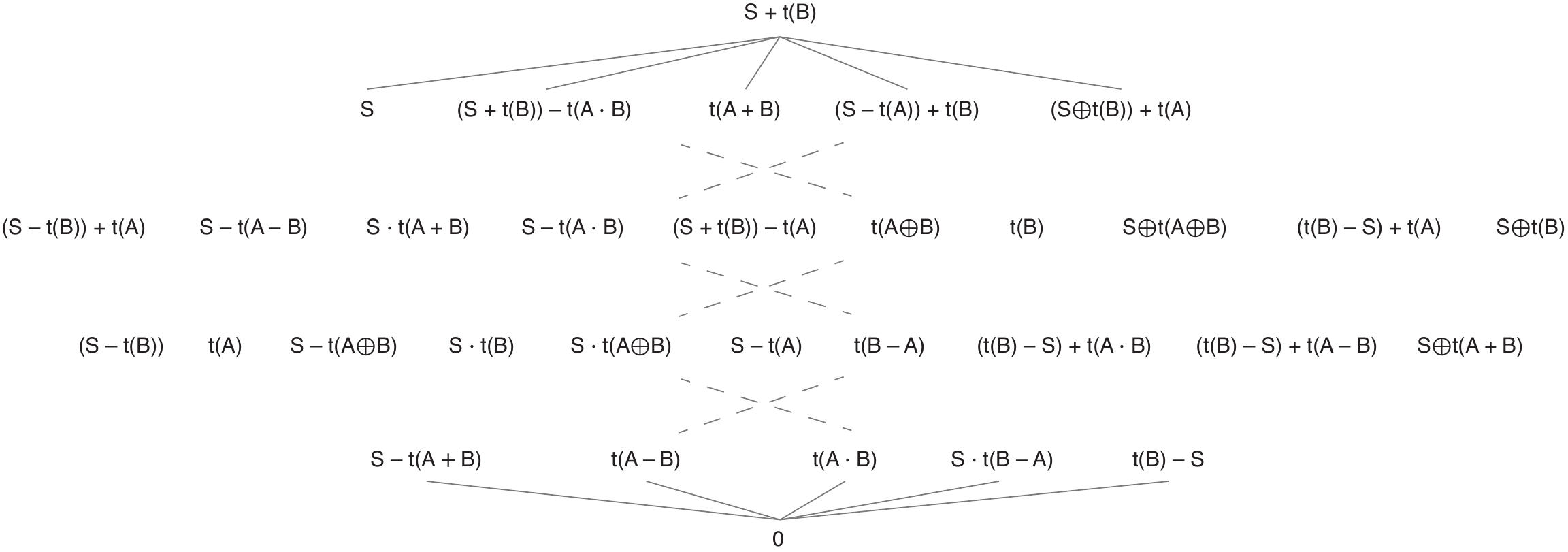}
\caption{Lattice diagram of production formulas characterizing a one-step rule application. The upward lines connecting the elements in the lattice are omitted for visual clarity. Adapted from Krstic (2019).}
\label{Figure-3}
\end{figure}

The formula $S' = (S - t(A)) + t(B)$ corresponds to a mapping $h: S \rightarrow S'$, from a given shape $S$ to a new shape $S'$ that the formula generates, and is defined as

\vspace{12pt}
\centerline{$h(x) = (x - t(A)) + t(B)$} 
\vspace{12pt}

\noindent for every part $x$ of $S$; in this case, we have $h(S) = S'$.

In general, production formulas are formed by combining the shapes $S$, $t(A)$, and $t(B)$---the three shapes that participate virtually in any rule application---with the usual operations that determine algebras of shapes, namely, sum (+), difference (--), product ($\cdot$), and symmetric difference ($\oplus$). Krstic (2019) shows a host of formulas that describe a one-step rule application---altogether, they form the thirty-two element lattice shown in Figure 3. These formulas are different ``ways of seeing" or selectively monitoring a one-step rule application based on how the three shapes are combined in operations. It is straightforward to translate any of these formulas into a corresponding mapping form.

\subsection{Characterizing the suitable forms}

The lattice in Figure 3 provides a practical starting point for experimenting with alternative forms for the mapping h. But not all mappings suggested by these formulas have a form that is suitable for analyzing rule continuity. Some forms of mappings are never continuous, no matter what shapes or topologies are involved. And some other forms are inconsequential. The following results, characterize three specific forms of mappings that are excluded as candidates for analyzing rule continuity.

Let $S$ and $S'$ be any two shapes. Suppose $h: S \rightarrow S^{+}$ is a mapping from the shape $S$ to a part $S^{+}$ of $S'$ and assume that $\mathcal{T}$ and $\mathcal{T}'$ are the topologies for $S$ and $S'$, respectively. If $h(x) \neq 0$ for every part $x$ of $S$, then the mapping $h$ is not continuous.

\begin{proof}
Suppose $h$ is continuous. Then, the empty shape, which is an open part in $OS^{+}$, must have a preimage $h^{-1}(0)$ open in $OS$ by definition of continuity. But, by assumption, there is no part $x$ of $S$ for which $h(x) = 0$ (i.e., no part $x$ maps to the empty shape under $h$), and so the preimage of the empty shape must be undefined. Thus, $h$ cannot be continuous. 
\end{proof}

This first result builds on the observation that the empty shape is a point of discontinuity when a mapping's ``net result," or calculated output, is always a \emph{nonempty} shape. This is a consequence of how the preimage is defined for shapes. In particular, the preimage of a part under a mapping is equal to the largest shape whose image under the mapping is embedded in that part. If no such shape exists, then the preimage is undefined (see \hyperlink{supA}{Appendix A}). If a mapping's calculated output is always nonempty, it can never map a part of $S$ to the empty shape. This means the empty shape cannot have a preimage under that mapping---a requirement for continuity of $h$.

Consider, for example, the mapping $h: S \rightarrow (S - t(A)) + t(B)$ defined just above. If in a particular rule application the shape $t(B)$ is nonempty, then $h(x) \neq 0$ for any part $x$ of $S$. It follows that $h$ cannot be continuous no matter what topologies are assigned to the shapes $S$ and $S'$. This is an example of a mapping that takes account of parts \emph{added} to the shape $S$---an empty shape can never be obtained by adding nonempty parts. 

The lattice in Figure 3 contains other formulas that take account of ``added parts" and which face a similar discontinuity problem. Two examples are 

\vspace{12pt}
\centerline{$h: S \rightarrow (S - t(B)) + t(A)$ and $h: S \rightarrow (S \oplus t(B)) + t(A)$.} 
\vspace{12pt}

\noindent In general, a rule may add parts to a shape $S$ that are not necessarily ``new." Indeed, a rule may add parts that are already embedded in $S$ that will fuse in $S$ once the rule is applied. A characterization of the parts added in a rule application is given in \hyperref[supB]{Appendix B}. 

Now, suppose in a rule application the shape $t(B)$ is empty. Then, $h: S \rightarrow (S - t(A)) + t(B)$ reduces to the equivalent, albeit simpler, mapping $h: S \rightarrow S - t(A)$ defined as $h(x) = x - t(A)$ for any part $x$ of $S$. This mapping does not face the same discontinuity problem as the mappings mentioned above. In particular, there is at least one part $x$ of $S$ for which $h(x)$ = $0$, namely, the part $x$ = $t(A)$. This is an example of a mapping that takes account of parts that are \emph{erased} from $S$ or are \emph{left untouched}, and ignores any parts that may be ``outside of" or ``external to" $S$. Readers familiar with the first study on rule continuity in Stiny (1994) will be quick to recognize this mapping since it was exclusively used in all the examples worked out in the paper.

Mappings that add nonempty parts to $S$ are not the only ones facing a discontinuity problem related to the empty shape. In particular, we have the following corollary concerning a particular class of mappings called ``constant mappings." 

A \emph{constant mapping} $h: S \rightarrow y_0$ is a mapping that outputs the same part $y_0$ of a shape $S'$ for every part $x$ of $S$ it takes as an input. If the part $y_0$ is nonempty, then $h(x) \neq 0$ and it follows from the preceding result that $h$ is not continuous. Moreover, it follows trivially that the mapping $h: S \rightarrow 0$ that outputs the empty shape, is the only constant mapping between shapes that is continuous. But this mapping is practically vacuous in the context of rule continuity.

The lattice in Figure 3 contains many constant mappings. For example, these ones

\vspace{12pt}
\centerline{$h: S \rightarrow t(A)$, $h: S \rightarrow t(B)$, and $h: S \rightarrow t(A - B)$.} 
\vspace{12pt}

\noindent It is useful to have a simple practical condition for deciding if a given mapping $h$ has the empty shape as a point of discontinuity in a particular analysis. Such a condition can be obtained through the following intermediary result.

\begin{table}[t!]
\small\sf\centering
\caption{Suitable mapping forms for analyzing rule continuity.\label{T1}}
\begin{tabular}{lllll}
\toprule
\midrule
\vspace{4pt}
$S$\\
\vspace{4pt}
$S - t(A - B)$ & $S \cdot t(A + B)$ & $S - t(A \cdot B)$\\
\vspace{4pt}
$S - t(B)$ & $S - t(A \oplus B)$ & $S \cdot t(B)$ & $S \cdot t(A \oplus B)$ & $S - t(A)$\\
$S - t(A + B)$ & $S \cdot t(B - A)$\\
\bottomrule
\end{tabular}
\end{table}

Again, let $S$ and $S'$ be any two shapes and $h: S \rightarrow S^{+}$ a mapping from the shape $S$ to a part $S^{+}$ of $S'$. If $h(0) \neq 0$, then $h(x) \neq 0$ for every part $x$ of $S$.

\begin{proof}
Suppose, for contradiction, that $h(0) \neq 0$ but there is a nonempty part $x_0$ of $S$ for which $h(x_0) = 0$. Then, $0 \leq x_0$ implies $h(0) \geq h(x_0)$. But this violates the requirement that $h$ is order-preserving. Thus, there is no such part $x_0$ of $S$.
\end{proof}

\noindent Now, using this intermediary result in conjunction with the first result, we have that if $h(0) \neq 0$, then $h$ is not continuous. This condition is easy to check for any mapping that is given in closed form.

The proof uses the requirement that a mapping $h$ between shapes is an order-preserving mapping; for why this is the case, see \hyperlink{supA}{Appendix A}. Order preservation is another feature to look for when deciding between candidate forms of mappings. Examples of order-reversing mappings in Figure 3 are 

\vspace{12pt}
\centerline{$h: S \rightarrow t(B) - S$ \;and\; $h: S \rightarrow (t(B) - S) + t(A \cdot B)$.} 
\vspace{12pt}

Table 1 contains a subset of the formulas in Figure 3. This table provides suitable forms of mappings for analyzing rule continuity and excludes the three forms of mappings we described in this section: order-reversing mappings, constant mappings, and mappings that cannot output the empty shape.

\section{Examples}

The continuity of a computation is analyzed in terms of the rules applied to generate shapes sequentially. Suppose a computation produces a sequence of shapes $S_1$, ..., $S_n$, where $S_1$ is the starting shape. The computation is continuous for the topologies assigned to the shapes $S_1$, ..., $S_n$ whenever every rule application in this sequence is continuous. The following examples illustrate how to use the given definition of continuity and the associated technical devices.

\begin{table}[t!]
\small\sf\centering
\caption{Topologies formed for the computation in Figure 1(b) when just the parts $t(A)$ are open.}
\begin{tabular}{l|lll}
\toprule
Shape & Topology\\
\midrule
\includegraphics[scale=0.70]{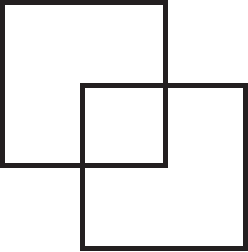} & \includegraphics[scale=0.70]{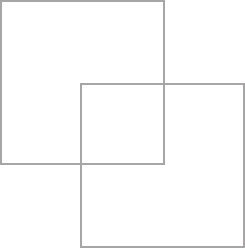} & \includegraphics[scale=0.70]{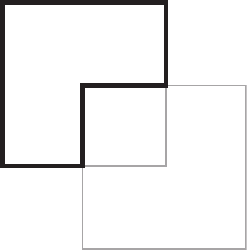} & \includegraphics[scale=0.70]{Figures/Tables/FigureEX1-T1-00.pdf}\\
\includegraphics[scale=0.70]{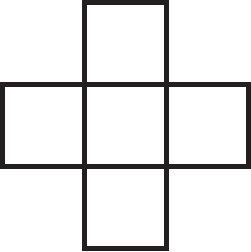} & \includegraphics[scale=0.70]{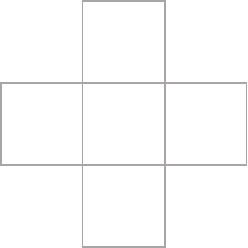} & \includegraphics[scale=0.70]{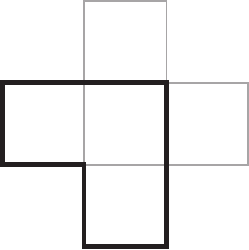} & \includegraphics[scale=0.70]{Figures/Tables/FigureEX1-T2-00.pdf}\\
\includegraphics[scale=0.70]{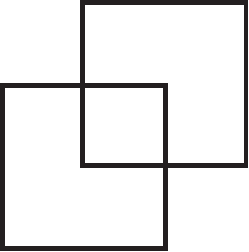} & \includegraphics[scale=0.70]{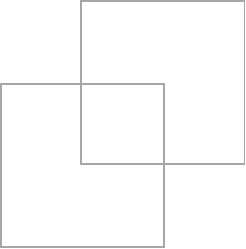} & \includegraphics[scale=0.70]{Figures/Tables/FigureEX1-T3-00.pdf}\\
\bottomrule
\end{tabular}
\end{table}

\subsection{Continuity with a single mapping}

The first example goes back to the two-step computation we used in the introductory section of this paper. Suppose the mapping $h_1: S \rightarrow S - t(A)$ defined in the usual way describes the two rule applications. Continuity analysis moves backward from the second rule application to the first one, to assign topologies to the three shapes that make each rule application continuous with respect to the selected mapping.

A rule application is continuous when the two conditions given in the definition of continuity are satisfied. To satisfy the first condition, the topologies in Table 2 are defined minimally to keep the shape $t(A)$ open in each step. The third shape in Table 2 has the indiscrete (two-part) topology because no rule is applied to divide it into parts. The second condition requires that $h$ yields a homomorphism for the topologies assigned to the three shapes. This is decided in a recursive fashion in the following way. 

The preimages of the open parts in the third topology must be open in the second topology. If not, the second topology must be refined to include them. Then, the preimages of the open parts in the second topology---including any new parts obtained after the refinement in the previous step---must be open in the first topology. If not, the first topology must be refined to include them.

\begin{table}[t!]
\small\sf\centering
\caption{Preimages for the open parts in the second topology in Table 2.}
\begin{tabular}{l|l}
\toprule
Open part $D$ & $h^{-1}(D)$\\
\midrule
\includegraphics[scale=0.70]{Figures/Tables/FigureEX1-T2-00.pdf} & \includegraphics[scale=0.70]{Figures/Tables/FigureEX1-T1-00.pdf}\\ \includegraphics[scale=0.70]{Figures/Tables/FigureEX1-T2-01.pdf} & \includegraphics[scale=0.70]{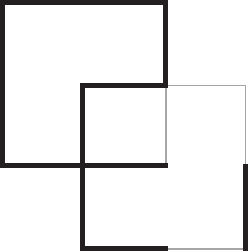}\\
\includegraphics[scale=0.70]{Figures/Tables/FigureEX1-T2-02.pdf} & \includegraphics[scale=0.70]{Figures/Tables/FigureEX1-T1-01.pdf}\\
\bottomrule
\end{tabular}
\end{table}

In more detail, the third topology has two open parts and their preimages under $h_1$ are already open in the second topology. So the second topology needs no updates to make the second rule application continuous. Next, the preimages under $h_1$ of the open parts in the second topology are shown in Table 3. The first rule application is not continuous because this shape

\begin{figure}[h!]
\centering
\includegraphics{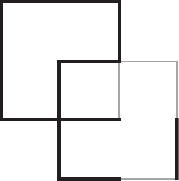}
\label{Figure-5-nocaption}
\end{figure}

\noindent is not an open part in the first topology. The first topology must be refined in this way

\begin{figure}[h!]
\centering
\includegraphics{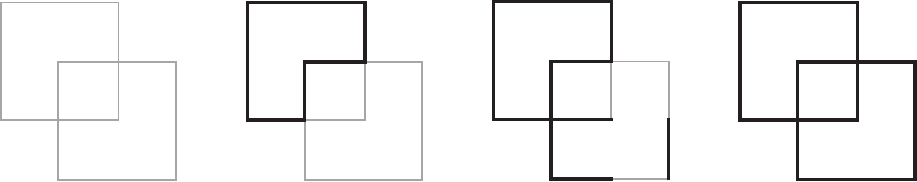}
\label{Figure-6-nocaption}
\end{figure}

\noindent to include this missing part. This refined topology for the first shape and the topologies for the second and third shapes given in Table 2 are the minimum required to make the two-step computation continuous with respect to the mapping $h_1$.

\begin{table}[b!]
\small\sf\centering
\caption{Continuous topologies for the shapes in Table 2, formed under the mapping $h_1: S \rightarrow S - t(A)$ when the parts $t(A)$ and $S - t(A)$ are open.}
\begin{tabular}{l|llll|l}
\toprule
Shape & Topology & & & & Mapping\\
\midrule
\includegraphics[scale=0.70]{Figures/Tables/FigureEX1-T1-00.pdf} & \includegraphics[scale=0.70]{Figures/Tables/FigureEX1-T1-02.pdf} & \includegraphics[scale=0.70]{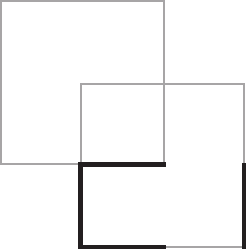} & \includegraphics[scale=0.70]{Figures/Tables/FigureEX1-T1-01.pdf} & \includegraphics[scale=0.70]{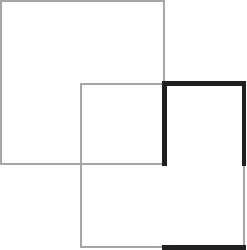} & \\
& \includegraphics[scale=0.70]{Figures/Tables/FigureEX1-T1-03.pdf} & \includegraphics[scale=0.70]{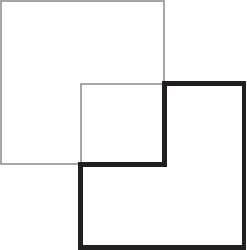} & \includegraphics[scale=0.70]{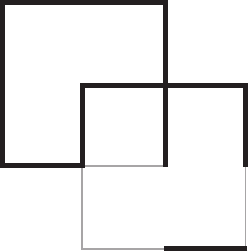} & \includegraphics[scale=0.70]{Figures/Tables/FigureEX1-T1-00.pdf} & $h_1: S \rightarrow S - t(A)$\\
\includegraphics[scale=0.70]{Figures/Tables/FigureEX1-T2-00.pdf} & \includegraphics[scale=0.70]{Figures/Tables/FigureEX1-T2-02.pdf} &
\includegraphics[scale=0.70]{Figures/Tables/FigureEX1-T2-01.pdf} & 
\includegraphics[scale=0.70]{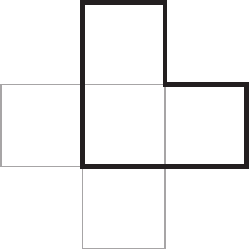} & \includegraphics[scale=0.70]{Figures/Tables/FigureEX1-T2-00.pdf} & \\
\includegraphics[scale=0.70]{Figures/Tables/FigureEX1-T3-00.pdf} & \includegraphics[scale=0.70]{Figures/Tables/FigureEX1-T3-01.pdf} & \includegraphics[scale=0.70]{Figures/Tables/FigureEX1-T3-00.pdf} & & &\\
\bottomrule
\end{tabular}
\end{table}

More elaborate topologies can be crafted by recognizing additional parts in the shapes participating in a rule application. For example, apart from keeping the part $t(A)$ open, one may additionally keep its complement open, the shape $S - t(A)$. In this case, the computation in Figure 1(b) becomes continuous with respect to the same mapping but the topologies that make this happen are those in Table 4 (these topologies form Boolean algebras). The shapes $t(A)$ and $S - t(A)$ are still not the only options. Different shapes formed in combinations of $S$, $S'$, $t(A)$, and $t(B)$ are readily observed in rule applications according to interest and purpose. The same computation can become continuous in alternative ways, depending on what parts one chooses to observe in the action of rules so long as the two conditions for continuity are met. Analysis of rule continuity is an observational process.

\subsection{Continuity with multiple mappings}

In this second example, we show how to use multiple mappings in an analysis. The effects of chaining different mappings in the same analysis are telling in longer computations.

\begin{figure}[t!]
\centering
\includegraphics{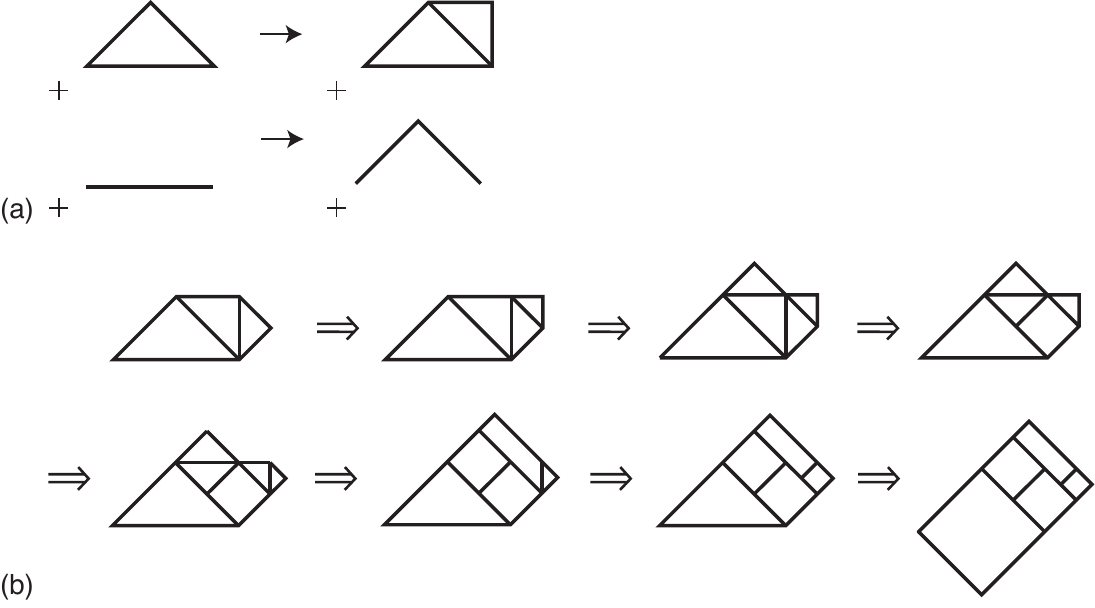}
\caption{(a) A shape grammar with two rules and (b) an example computation that uses these two rules. The left side of the second rule in (a) is meant to match only the maximal elements of a shape; this makes it a determinate rule (this approach appears in Stiny (1975)).}
\label{Figure-4}
\end{figure}

Consider the two rules in Figure 4(a) and the way they're applied in the computation in Figure 4(b). Suppose the first two rule applications correspond to the mapping $h_2: S \rightarrow S - t(B)$, defined as $h_2(x) = x - t(B)$ for every part $x$ of $S$. Then, the third rule application corresponds to the mapping $h_3: S \rightarrow S - t(A \oplus B)$, defined as $h_3(x) = x - t(A \oplus B)$ for every part $x$ of $S$. The fourth rule application corresponds to the mapping $h_2$, and finally, the fifth and the rest of the rule applications correspond to the mapping $h_3$. 

To keep the topologies simple and small in size, suppose only the mandatory part $t(A)$ is kept open in each step of the computation. The topologies that make the computation continuous with respect to both $h_2$ and $h_3$ are shown in Table 5.

In this analysis, we alternate between the two given mappings $h_2$ and $h_3$ for two main reasons. First, to illustrate the flexibility and generality of the proposed approach to continuity analysis, and the ease with which one can define and use alternative mappings to analyze different stages (time frames) of the same computation. Second, to highlight a technical point in the definition of continuity. In particular, recall that if $h$ is some mapping that describes a rule application $S \Rightarrow S'$, then $h$ maps the shape $S$ into some part $S^{+}$ of the shape $S'$, that is to say, it must be $h(S) = S^{+} \leq S'$. But in the third rule application in Table 5, we cannot use the mapping $h_2$ because it is not a description of the rule application $S_3 \Rightarrow S_4$ such that $h_2(S_3) \leq S_4$, that is, the image $h_2(S_3)$ is not a part of $S_4$. The mapping $h_3$ is used instead. The same thing happens in the sixth rule application in the table.

The topologies formed with respect to $h_3$ in the last steps of the computation propagate their open parts backward until the first shape, intermixing along the way with the structures formed with respect to $h_2$. Looking ahead, the structures give the impression of a seamless forward computation, as if the topologies of the shapes in earlier steps anticipate the progression of the rule applications in subsequent steps and the structures that emerge as a consequence of their action.

\begin{table}[t!]
\small\sf\centering
\caption{Continuous topologies formed under the mappings $h_2: S \rightarrow S - t(B)$ and $h_3: S \rightarrow S - t(A \oplus B)$, when the shapes $t(A)$ are open.}
\begin{tabular}{l|lllll|l}
\toprule
Shape & Topology & & & & & Mapping\\
\midrule
\includegraphics[scale=0.50]{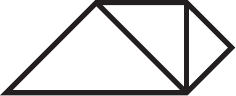} & \includegraphics[scale=0.50]{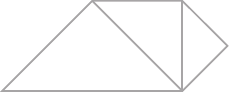} & \includegraphics[scale=0.50]{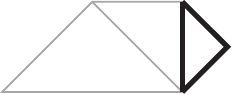} & \includegraphics[scale=0.50]{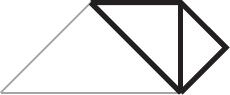} & \includegraphics[scale=0.50]{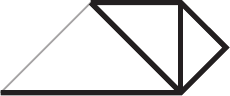} & \includegraphics[scale=0.50]{Figures/Tables/FigureEX2-T1-00.pdf} & $h_2: S \rightarrow S - t(B)$\\
\includegraphics[scale=0.50]{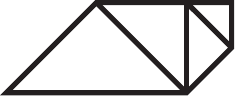} & \includegraphics[scale=0.50]{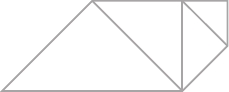} & \includegraphics[scale=0.50]{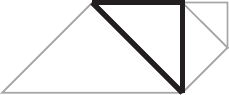} & \includegraphics[scale=0.50]{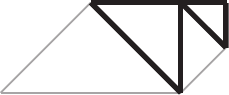} & \includegraphics[scale=0.50]{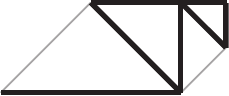} & \includegraphics[scale=0.50]{Figures/Tables/FigureEX2-T2-00.pdf} & \\
\includegraphics[scale=0.50]{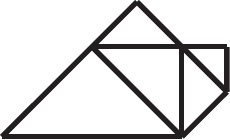} & \includegraphics[scale=0.50]{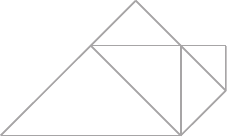} & \includegraphics[scale=0.50]{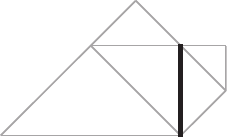} & \includegraphics[scale=0.50]{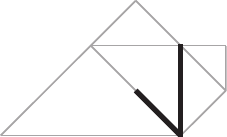} & \includegraphics[scale=0.50]{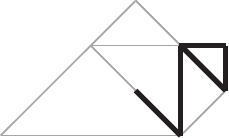} & \includegraphics[scale=0.50]{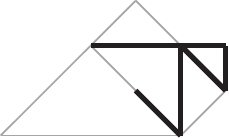} & \\
& \includegraphics[scale=0.50]{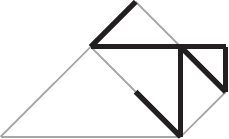} & \includegraphics[scale=0.50]{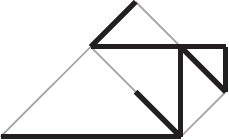} & \includegraphics[scale=0.50]{Figures/Tables/FigureEX2-T3-00.pdf} & & & $h_3: S \rightarrow S - t(A \oplus B)$\\
\includegraphics[scale=0.50]{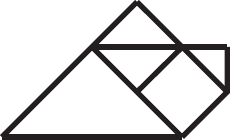} & \includegraphics[scale=0.50]{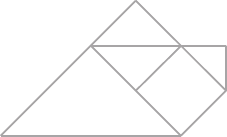} & \includegraphics[scale=0.50]{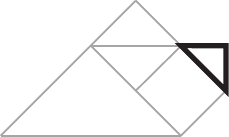} & \includegraphics[scale=0.50]{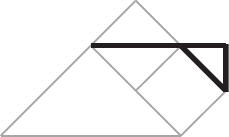} & \includegraphics[scale=0.50]{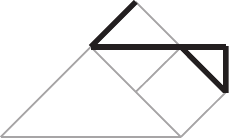} & \includegraphics[scale=0.50]{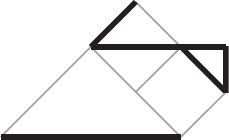} & \\
& \includegraphics[scale=0.50]{Figures/Tables/FigureEX2-T4-00.pdf} & & & & & $h_2: S \rightarrow S - t(B)$\\
\includegraphics[scale=0.50]{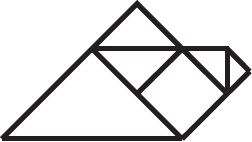} & \includegraphics[scale=0.50]{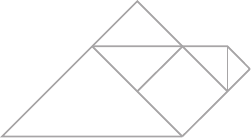} & \includegraphics[scale=0.50]{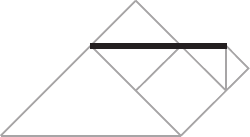} & \includegraphics[scale=0.50]{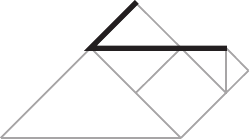} & \includegraphics[scale=0.50]{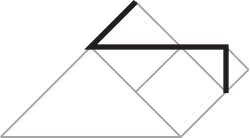} & \includegraphics[scale=0.50]{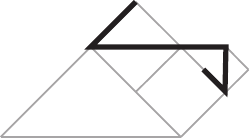} & \\
& \includegraphics[scale=0.50]{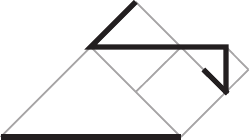} & \includegraphics[scale=0.50]{Figures/Tables/FigureEX2-T5-00.pdf} & & & & $h_3: S \rightarrow S - t(A \oplus B)$\\
\includegraphics[scale=0.50]{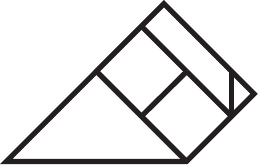} & \includegraphics[scale=0.50]{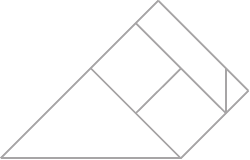} & \includegraphics[scale=0.50]{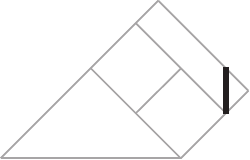} & \includegraphics[scale=0.50]{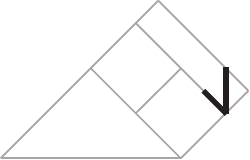} & \includegraphics[scale=0.50]{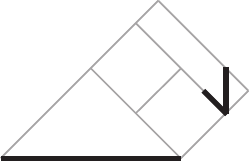} & \includegraphics[scale=0.50]{Figures/Tables/FigureEX2-T6-00.pdf} &\\
\includegraphics[scale=0.50]{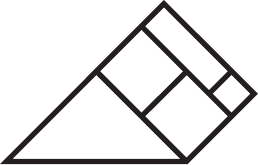} & \includegraphics[scale=0.50]{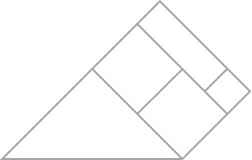} & \includegraphics[scale=0.50]{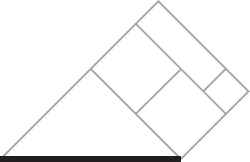} & \includegraphics[scale=0.50]{Figures/Tables/FigureEX2-T7-00.pdf} & & &\\
\includegraphics[scale=0.50]{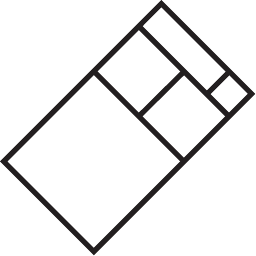} & \includegraphics[scale=0.50]{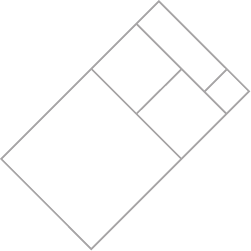} & \includegraphics[scale=0.50]{Figures/Tables/FigureEX2-T8-00.pdf} & & & &\\
\bottomrule
\end{tabular}
\end{table}

\section{Computability of continuous topologies}

Suppose we have recorded the rule applications in a computation for a certain number of steps. Continuity analysis includes two main computational tasks:

\begin{enumerate}
\item[(1)] The generation and refinement of shape topologies by given parts, and

\item[(2)] The calculation of preimages of parts under given mapping(s).
\end{enumerate}

\noindent The two tasks are carried out for each rule application, starting from the rule application recorded last until the first one.

For task (1), since the topologies we deal with are finite their computability is guaranteed. A topology is generated or refined by given parts by (re)calculating finitary sums and products---a recursive procedure for this is given in Haridis (2020a).

For task (2), the preimages can be calculated with the help of closure equations, which were first introduced in Stiny (1994). Here, we revise the closure equations in terms of preimages of mappings. If a mapping $h$ between two shapes $S$ and $S'$ is continuous for some topologies assigned to the two shapes, then the following ``closure under image" inequality 


\vspace{12pt}
\centerline{$h(\overline{x}) \leq \overline{h(x)}$} 
\vspace{12pt}

\noindent always holds for every part $x$ of $S$ (Haridis, 2020a: p. 225). If $x$ is open, then x is equal to its closure $\overline{x}$. 

Any open part $D$ of the shape $S'$ for which we wish to compute its preimage must be the shape that is ultimately formed in the right side of this inequality. In other words, the preimage $h^{-1}(D)$ we wish to compute should be the largest shape that is open in the topology for the shape $S$ satisfying this inequality

\vspace{12pt}
\centerline{$h(h^{-1}(D)) \leq D$.} 
\vspace{12pt}

\begin{figure}[t!]
\centering
\includegraphics{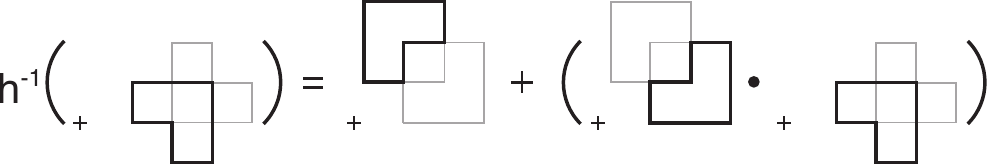}
\caption{An example calculation of the preimage of an open part using the closure equation $h^{-1}(D) = t(A) + (S - t(A))\cdot D$. The open part is taken from the topologies in Table 2 (or 4).}
\label{Figure-5}
\end{figure}

Now, the actual calculation of the preimage shape $h^{-1}(D)$ depends on the closed form of the particular mapping $h$ that is chosen in an analysis. For example, for the mapping $h_1: S \rightarrow S - t(A)$ defined as $h_1(x) = x - t(A)$ for any part $x$ of $S$, the inequality becomes

\vspace{12pt}
\centerline{$h_1^{-1}(D) - t(A) \leq D$.} 
\vspace{12pt}

\noindent By constraining the right side in terms of the shape $S - t(A)$---that is the shape that is guaranteed to stay the same in both $S$ and $S'$ when the rule is applied---we obtain the equation

\vspace{12pt}
\centerline{$h_1^{-1}(D) - t(A) = (S - t(A))\cdot D$} 
\vspace{12pt}

\noindent which can be solved for the desired preimage shape $h_1^{-1}(D)$. This equation can be used to verify the preimages computed in the first analysis in the section \emph{Examples}, and especially Tables 3 and 4. Figure 5 shows one application of this equation.

As another example, for the two mappings $h_2: S \rightarrow S - t(B)$ and $h_3: S \rightarrow S - t(A \oplus B)$ we used in the second analysis in the section \emph{Examples}, the inequality is satisfied in this way

\vspace{12pt}
\centerline{$h_2^{-1}(D) - S \cdot t(B) = (S - t(B))\cdot D$ \;and\; $h_3^{-1}(D) - S \cdot t(A \oplus B) = (S - t(A \oplus B))\cdot D$.}
\vspace{12pt}

\noindent The products $S \cdot t(B)$ and $S \cdot t(A \oplus B)$ are necessary because the shapes $t(B)$ and $t(A \oplus B)$ may have parts that are not parts of the shape $S$. Both equations can be used to verify the topologies in Table 5.

Analogous equations can be defined for other suitable  mappings that are given in closed form, so long as the second of the above inequalities is satisfied. These equations can be used for individual applications of rules, but they can be used recursively, too, beginning with the topology of the final shape in a recorded computation and working backward to the initial shape.

\begin{table}[t!]
\small\sf\centering
\caption{Two examples of topologies for the last two shapes in Table 4. The numbers in the second topology indicate the basis elements.}
\begin{tabular}{l|lllll}
\toprule
Shape & Topology & & & &\\
\midrule
\includegraphics[scale=0.70]{Figures/Tables/FigureEX2-T7-00.pdf} & \includegraphics[scale=0.70]{Figures/Tables/FigureEX2-T7-02.pdf} & \includegraphics[scale=0.70]{Figures/Tables/FigureEX2-T7-01.pdf} & \includegraphics[scale=0.70]{Figures/Tables/FigureEX2-T7-00.pdf} & & \\
\includegraphics[scale=0.70]{Figures/Tables/FigureEX2-T8-00.pdf} & 1\includegraphics[scale=0.70]{Figures/Tables/FigureEX2-T8-01.pdf} & 2\includegraphics[scale=0.70]{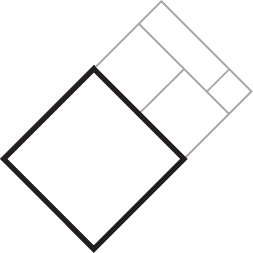} & 3\includegraphics[scale=0.70]{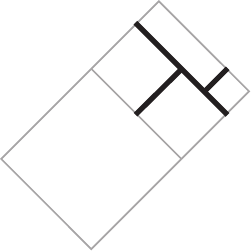} & 4\includegraphics[scale=0.70]{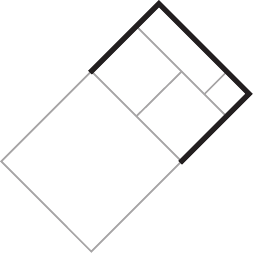} & \includegraphics[scale=0.70]{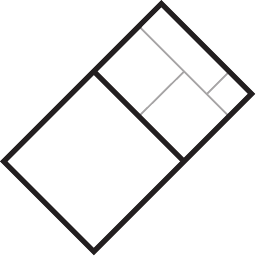} \\
& \includegraphics[scale=0.70]{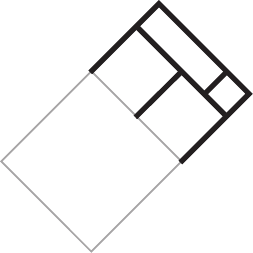} & \includegraphics[scale=0.70]{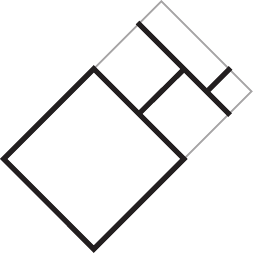} & \includegraphics[scale=0.70]{Figures/Tables/FigureEX2-T8-00.pdf}\\
\bottomrule
\end{tabular}
\end{table}

A topological concept that is very useful for calculating preimages in task (2) is the concept of a \emph{reduced basis} (Haridis 2020a: 214-215). Every topology for a shape has a unique minimal set of parts that describes the topology, i.e., generates all its open parts. This set is called a reduced basis and the parts in it are called basis elements. 

When a topology is large, calculating the preimages for each and every open part may be a time inefficient task. One way to avoid this is to use the reduced basis of the topology in place of the topology itself---the reduced basis acts as a representation or data structure for the topology. 

When we analyze continuity for a rule application $S \Rightarrow S'$, we can calculate the preimages only for the basis elements of the topology for $S'$. If we then use these calculated preimages to refine the topology for $S$ (i.e., re-evaluate sums and products), the refinement will automatically account for the preimages of all other open parts in the topology for $S'$, without us having to actually calculate them one by one.

The reduced basis is computationally beneficial when its size is smaller than the topology it describes. But this is not the case at all times. For example, in each topology in Table 5, the set of basis elements is equal to the topology itself; no topology contains open parts that sum to other open parts in the same topology. On the other hand, consider the hypothetical scenario in Table 6 where the two shapes participating in the last rule application in Table 5 acquire the two given topologies. Suppose $S \Rightarrow S'$ denotes this rule application. To establish continuity, the topology for $S$ would have to be refined in terms of the preimages of the open parts in the topology for $S'$ under some mapping. But instead of calculating the preimages for all eight open parts one can achieve the desired result by calculating the preimages just for the basis elements---there are four of them in total, labelled 1 through 4.

\section{Discussion: continuity in computational processes in design}

The technical tools presented in this paper can be used to analyze rule continuity in any computation that can be simulated by shape grammars. This includes the unrestricted kind of computations, in which shapes are unanalyzed, emergent parts can be recognized freely, and rules can be added (or removed) even as a computation unfolds. And it also includes computations of a more conventional kind, in which symbol structures are used as surrogates for shapes (representations), to prefigure what the rules must be and the most they can do in the course of a computation. Shape grammars can simulate computations of the latter kind with shapes that have points for basic elements (shapes in an algebra $U_0$) or, comparably, with shapes defined in terms of a vocabulary of combinatorial units and spatial relations.

The first of these representational techniques is especially clear in computations with parametric shapes (as in the computations we often see in the areas of design optimization and BIM) in which shapes are represented with points and the rules assign values in predefined numerical ranges to one or more preselected subsets of these points. The underlying representations of shapes work cooperatively with rules to fix precisely the parts (i.e., subsets of points) that can be recognized and changed in shapes in each state of a computation. Such a computation is inherently continuous, as the example demonstration in \hyperref[supC]{Appendix C} shows.

The second representational technique is employed when it is apparent how the shapes will be divided into parts in terms of a definite kit of parts. In this case, continuity can be guaranteed from the start simply by assigning a discrete topology to each shape in which all possible parts and their combinations are open (for discrete topologies and other topological structures on shapes made with point-like figures, see Haridis (2020b)). This approach has little explanatory value, in the sense that it tells nothing about what rules distinguish and change in shapes (``see and do") as a computation unfolds. Nevertheless, this approach can be pursued in computations where vocabularies of parts and spatial relations combining them are given in advance.

In practical applications, it is often the case that the decisions made about the end goals of a computation postulate particular shape descriptions. If one has already made a decision on what the shapes one works with should look like, and one is willing to forgo further exploration, then it is reasonable to pursue techniques that would allow for the analysis of these shapes into compatible structural descriptions (for example, many of the techniques we often see in computer graphics and shape analysis have this purpose, that is, to define geometric data structures that make seemingly different shapes be structurally compatible with one another). But this approach shows no promise when one is at the stage of creative exploration, where decisions about how shapes should look like and how they should be structured for a particular design problem are still in a state of uncertainty and change. Having first applied rules of computation in terms of embedding to create possibilities, one can inquiry afterwards how to go about structuring those possibilities into descriptions that do not contradict each other (that is, in relation to how the rules were applied to create them). An entirely natural way of proceeding for the analysis of the events involved in any creative activity.


%


\begin{dci}
The authors declare that there is no conflict of interest.
\end{dci}

\section*{Acknowledgement(s)}

This paper is the Authors version of a journal article published in \emph{Environment and Planning B: Urban Analytics and City Science} (SAGE), available at: \url{https://journals.sagepub.com/doi/10.1177/23998083211044734}{}. \textbf{Cite as}:

Haridis A. and Stiny G. (2022) Analysis of shape grammars: continuity of rules. \emph{Environment and Planning B: Urban Analytics and City Science}. doi:10.1177/23998083211044734

\newpage 

\begin{sm}
\textbf{A. Mappings between shapes}
\label{supA}

A one-step rule application $S \Rightarrow S'$ can be described by a \emph{mapping}

\vspace{12pt}
\centerline{$h: S \rightarrow S^{+}$} 
\vspace{12pt}

\noindent from the parts of the shape $S$ to a part $S^{+}$ of a new shape $S'$. Depending on the form of $h$, the shape $S^{+}$ may be equal to the full shape $S'$ or to a proper part of $S'$. If $x$ is a part of $S$, then $h(x)$ is the \emph{image} of $x$ under the mapping $h$---a unique shape that is part of $S^{+}$. Mappings provide a way of describing in different ways how the parts of the shape $S$ change or transform into parts of the shape $S'$ in a rule application.

It is well known that the set of all the parts of a shape $S$ forms a partially ordered set (poset) with a relation ``$\leq$", defined by $x \leq y$ for two parts $x, y$ of $S$, just when $x + y = y$, or equivalently $x \cdot y = x$. The relation ``$\leq$" is a part relation defined in terms of embedding (Stiny, 1975; 2006). Thus, a mapping between two shapes can be understood as a mapping between their respective posets. 

In this paper, the focus is on mappings that preserve the part relation. The mapping $h$ is order-preserving whenever for all parts $x, y$ of $S$ with $x \leq y$, we also have $h(x) \leq h(y)$ in $S^{+}$. By making this assumption, we obtain a highly desirable property: the \emph{image operation} under $h$ preserves the embedding order of parts. Throughout this paper, by ``$h$ is a mapping between shapes" it is always meant that $h$ is order-preserving.

It is possible to define a mapping between two shapes that is order-reversing. For example, this rule application 

\begin{figure}[h!]
\centering
\includegraphics{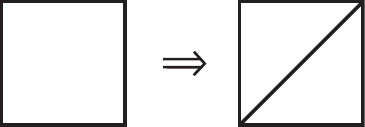}
\label{Figure-S1-nocaption}
\end{figure}

\noindent can be described with both of these mappings

\vspace{12pt}
\centerline{$h_1: S \rightarrow S + $ \includegraphics[height=0.4in,valign=m]{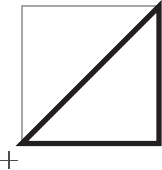} \;\;and\;\; $h_2: S \rightarrow $ \includegraphics[height=0.4in,valign=m]{Figures/Figure-S2-nocaption.pdf} $- S$} 
\vspace{12pt}

\noindent defined as $h_1(x) = x +$\includegraphics[height=0.4in,valign=m]{Figures/Figure-S2-nocaption.pdf} and $h_2(x) = $ \includegraphics[height=0.4in,valign=m]{Figures/Figure-S2-nocaption.pdf} $- x$. The first mapping is order-preserving while the second mapping is order-reversing. Order-preserving mappings have an appealing behavior in general, and are particularly useful for analyzing continuity of rules. Order-reversing mappings are more pedagogical in nature, and have no particularly useful role in analyzing continuity of rules.

If $y$ is a part of the shape $S^{+}$, denote by $h^{-1}(y)$ the largest part of $S$ whose image under $h$ is embedded in $y$ (Haridis, 2020a). The shape $h^{-1}(y)$ is called the \emph{inverse image} or \emph{preimage} of $y$. The shape $h^{-1}(y)$ is always defined when the part $y$ has a preimage under the given mapping $h$. On the other hand, if $y$ does not have a preimage in $S$ (i.e., if no part maps to it under $h$), then the preimage $h^{-1}(y)$ is said to be undefined. This is in contrast to what happens in set theory where, in the analogous scenario of a subset that does not have a preimage, the preimage of that subset is said to be equal to the empty set. It is not possible to replicate this in the case of shapes because the empty set is not the same object as the empty shape---they are categorically two different objects. For more information, see Haridis (2020a: 224--225). This possibility of having undefined preimages for certain parts of a shape, restricts the mappings that are suitable for continuity analysis as we discuss in the section \emph{How to choose mappings}.

\newpage

\textbf{B. Parts that a rule adds}
\label{supB}

The formula $(S - t(A)) + t(B)$ computes a new shape $S'$ in two basic stages, corresponding to two different actions. The part $S - t(A)$ is obtained after erasing the matched piece $t(A)$ from $S$, and the shape $t(B)$ is added to or equivalently drawn on top of $S - t(A)$, to replace $t(A)$. In general, assuming that $B$ is nonempty, the shape $t(B)$ can: add a part it already shares with $S$ back to $S - t(A)$; add a part that is completely new, and not already shared with $S$; or do both. The following computation shows the latter scenario.

Apply the additive rule

\begin{figure}[h!]
\centering
\includegraphics{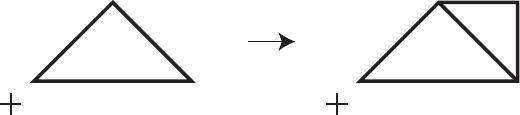}
\label{Figure-S3-nocaption}
\end{figure}

\noindent to the smaller triangle of the shape in the right-hand side, in this way

\begin{figure}[h!]
\centering
\includegraphics{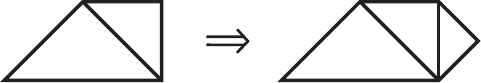}
\label{Figure-S4-nocaption}
\end{figure}

\noindent The shape $S - t(A)$, and the partitions of the shape $t(B)$ into $S \cdot t(B)$, which is the piece shared with $S$, and $t(B) - S$, which is the newly added piece, are these ones

\begin{figure}[h!]
\centering
\includegraphics{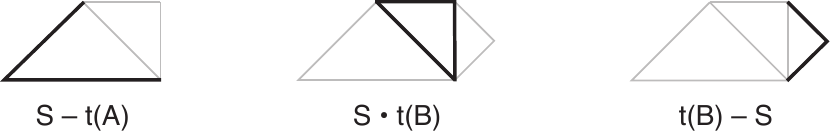}
\label{Figure-S5-nocaption}
\end{figure}

There are other ways to take account of parts added to a shape $S$ in a rule application, but these added parts need not necessarily correspond to the shape $t(B)$. For example, a formula such as $(S - t(B)) + t(A)$ erases the parts of $t(B)$---both the ones shared with $S$ and the ones that are new, if any---but adds the matched shape $t(A)$ back to $S$. In many cases, the result of this calculation may actually be equivalent to applying an identity rule or even an erasing rule even though it is obtained through addition. 

\newpage

\textbf{C. Continuity in computations with parametric shapes}
\label{supC}

\begin{figure}[t!]
\centering
\includegraphics[scale=0.95]{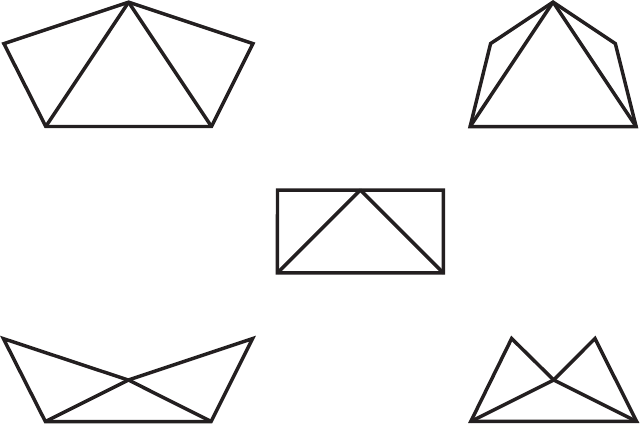}
\label{Figure-S6}
\caption{Designs generated by continuous variation of the same set of numerical parameters.}
\end{figure}

In computations where shapes are controlled indirectly by a fixed set of variables/parameters and are changed from one state of the computation to the other by assigning values to those variables, the rule applications involved are inherently continuous. Examples of designs generated in this way are shown in Figure 6.

This initial shape

\begin{figure}[h]
\centering
\includegraphics{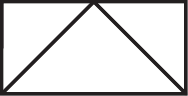}
\label{Figure-S7-nocaption}
\end{figure}

\noindent can be decomposed into seven line segments that are connected with five points in this way

\begin{figure}[h]
\centering
\includegraphics{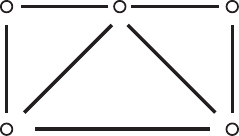}
\label{Figure-S8-nocaption}
\end{figure}

\noindent Then the initial shape can be parameterized in terms of this decomposition by taking advantage of its global symmetry as shown in the following figure

\begin{figure}[h]
\centering
\includegraphics{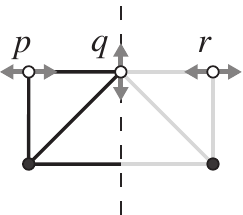}
\label{Figure-S9-nocaption}
\end{figure}

\noindent In particular, three points, namely $p$, $q$, and $r$, function as open terms (variables) and the rest of the points are fixed in place. The points $p$ and $r$ are allowed to move horizontally while the point $q$ is allowed to move vertically. The points $p$ and $r$ must take values in a symmetric fashion about the vertical axis passing through the point $q$ (i.e., if one's position changes, the other's position changes, too, in the opposite direction, but in equal magnitude). These constraints can be hard-coded into a schema $x \rightarrow y$

\begin{figure}[h]
\centering
\includegraphics{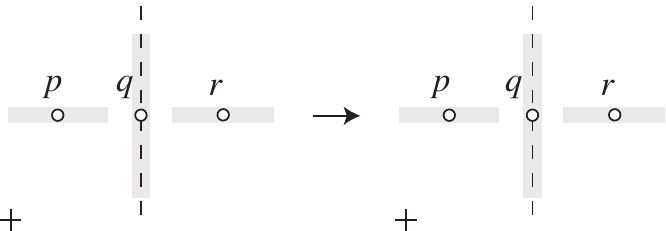}
\label{Figure-S10-nocaption}
\end{figure}

\vspace{1.5in}

\noindent made with the parameterized labelled shapes $x$ and $y$. The grey rectangles represent graphically the ranges of values that the three open terms are allowed to take.

When specific values are assigned to the open terms by a function, or ``rule of assignment," $g$ to determine exactly the shapes $x$ and $y$, a new rule $g(x) \rightarrow g(y)$ is defined. This rule applies if there is a transformation $t$ that matches $g(x)$ to a part (in this case, a subset) of a given shape. In this example, the rules apply to a subset that contains the open terms $p$, $q$, and $r$, in this way

\begin{figure}[h]
\centering
\includegraphics{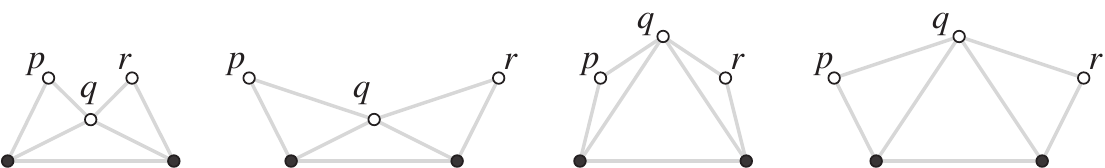}
\label{Figure-S11-nocaption}
\end{figure}

\noindent Because the rules apply only to sets of points, the lines (displayed in grey) play no role in the computation. The lines function only as ``links" between the points, to preserve the connectivity of the initial shape given by its decomposition. While in Figure 6 it appears as if the designs are generated by computing with lines, the actual computation happens with points.

The part matched in every rule application is $t(g(x))$ consisting of the three points $p$, $q$, and $r$ for a certain assignment of values by the function $g$. Topologies are defined for each of the shapes by keeping the subset of points $t(g(x))$ open. It is left as an exercise to verify that by keeping just this part open, the rule applications become automatically continuous (in fact, the rule applications are still continuous when the complement $S - t(g(x))$ is kept open, in addition to $t(g(x))$ itself).

Any shape made with line segments (a drawing in the plane) is a member of an algebra $U_1$. The representation of the shape in terms of a set of open and fixed terms is a shape made with finitely many points in an algebra $U_0$ augmented with labels (often called $V_0$). In a parametric computation, a shape (i = 1) is controlled by its underlying representation (i = 0) and only specified parts (point-sets) of this representation can be changed or transformed by rules. Structural discontinuities cannot occur in computations of this kind because the representations anticipate explicitly the parts that the rules can change in each state of a computation.

\end{sm}


\end{document}